\pdfoutput=1
\documentclass[12pt,a4paper]{article}
\usepackage{xcolor}
\usepackage{enumerate}
\usepackage{framed}
\usepackage{mdframed}
\usepackage{amsmath}
\usepackage{amsfonts}
\usepackage{mathrsfs}
\usepackage{amssymb}
\usepackage{graphicx, rotating}
\usepackage{epsfig}
\usepackage{latexsym}
\usepackage{graphicx}
\usepackage{color}
\usepackage{amsmath,bm,amssymb}
\usepackage{cite}
\usepackage{slashed}
\usepackage{hyperref}
\usepackage{epstopdf}
\usepackage[font=footnotesize,labelfont=]{caption}
\AppendGraphicsExtensions{.tif}
\hypersetup{colorlinks=true, citecolor=bluscuro, linkcolor=black, urlcolor=bluscuro}
\definecolor{rossos}{cmyk}{0,1,1,0.55}
\definecolor{bluscuro}{rgb}{0.15, 0.2, .85}
\definecolor{bluchiaro}{cmyk}{1,.3,0.,0.1}

\def\0{\vec{0}}

\def\d{{\rm d}}
\def\vx{{\vec{x}}}

\def\vk{{\vec{k}}}

\def\vq{{\vec{q}}}

\def\beq{\begin{equation}}
\def\eeq{\end{equation}}

\def\e{\zeta}

\newcommand{\bea}{\begin{eqnarray}}
\newcommand{\eea}{\end{eqnarray}}
\def\beqa{\begin{eqnarray}}

\def\eeqa{\end{eqnarray}}
\def\lsim{\mathrel{\rlap{\lower4pt\hbox{\hskip0.5pt$\sim$}}
    \raise1pt\hbox{$<$}}}         
\def\gsim{\mathrel{\rlap{\lower4pt\hbox{\hskip0.5pt$\sim$}}
    \raise1pt\hbox{$>$}}}         
\def\e{\zeta}

\def\d{{\rm d}}
\def\vx{{\vec{x}}}
\def\vy{{\vec{y}}}

\def\vk{{\vec{k}}}

\def\vq{{\vec{q}}}

\usepackage[normalem]{ulem}

\begin{document}
\def\thefootnote{\fnsymbol{footnote}}

\begin{center}
\Large{\textbf{Primordial Black Holes from Inflation and non-Gaussianity}} \\[0.5cm]
\end{center}
\vspace{0.5cm}

\begin{center}

\large{G. Franciolini$^{\rm a}$\footnote{gabriele.franciolini@unige.ch}, A. Kehagias$^{\rm b}$\footnote{kehagias@central.ntua.gr},  S. Matarrese$^{c,d,e,f}$\footnote{sabino.matarrese@pd.infn.it} and  A.~Riotto$^{\rm a}$\footnote{antonio.riotto@unige.ch}}
\\[0.5cm]

\small{
\textit{$^{\rm a}$Department of Theoretical Physics and Center for Astroparticle Physics (CAP) \\
24 quai E. Ansermet, CH-1211 Geneva 4, Switzerland}}

\small{
\textit{$^{\rm b}$Physics Division, National Technical University of Athens, 15780 Zografou Campus, Athens, Greece}}

\small{
\textit{$^{\rm c}$Dipartimento di Fisica e Astronomia ``G. Galilei'', Universit\`a degli Studi di Padova\\ via Marzolo 8, I-35131, Padova, Italy}}

\small{
\textit{$^{\rm d}$INFN, Sezione di Padova, via Marzolo 8, I-35131, Padova, Italy}}

\small{
\textit{$^{\rm e}$INAF-Osservatorio Astronomico di Padova, vicolo dell Osservatorio 5, I-35122 Padova, Italy}}

\small{
\textit{$^{\rm f}$Gran Sasso Science Institute,
via F. Crispi 7, I-67100, L'Aquila, Italy}}

\vspace{.2cm}


\vspace{.2cm}

\vspace{.2cm}

\end{center}

\vspace{.7cm}

\hrule \vspace{0.3cm}
\noindent \small{\textbf{Abstract}\\ 
Primordial black holes may owe their origin  to the  small-scale enhancement of the comoving curvature perturbation  generated during inflation.
Their mass fraction at formation  is markedly sensitive to  possible  non-Gaussianities in such large, but rare  fluctuations. We discuss a  path-integral formulation which provides the exact  mass fraction of primordial black holes at formation in the presence of non-Gaussianity. Through a couple of classes of models, one based on single-field inflation and the other on spectator fields,  we show  that restricting to a Gaussian statistics may  lead to severe  inaccuracies in the estimate of the  mass fraction as well as on the clustering properties of the primordial black holes.  }

\vspace{0.3cm}
\noindent
\hrule
\def\thefootnote{\arabic{footnote}}
\setcounter{footnote}{0}

%
%
%
%
%

%
%
%
%


 \def\vx{\vec{ x}} 
\def\vk{\vec{k}}
\def\vy{\vec{y}}

\numberwithin{equation}{section}

\def\la{~\mbox{\raisebox{-.6ex}{$\stackrel{<}{\sim}$}}~}
\def\ga{~\mbox{\raisebox{-.6ex}{$\stackrel{>}{\sim}$}}~}
\def\bq{\begin{quote}}
\def\eq{\end{quote}}
\def\PL{{ \it Phys. Lett.} }
\def\PRL{{\it Phys. Rev. Lett.} }
\def\NP{{\it Nucl. Phys.} }
\def\PR{{\it Phys. Rev.} }
\def\MPL{{\it Mod. Phys. Lett.} }
\def\IJMP{{\it Int. J. Mod .Phys.} }
\font\tinynk=cmr6 at 10truept
\newcommand{\be}{\begin{eqnarray}}
\newcommand{\ee}{\end{eqnarray}}
\newcommand{\n}{{\bf n}}
\newcommand{\arXiv}[2]{\href{http://arxiv.org/pdf/#1}{{\tt [#2/#1]}}}
\newcommand{\arXivold}[1]{\href{http://arxiv.org/pdf/#1}{{\tt [#1]}}}

\section{Introduction}
\noindent
With the recent detection of gravitational  waves originated by  two $\sim 30 M_\odot$ black holes \cite{ligo} the idea that Primordial Black Holes (PBHs)
might form a considerable fraction of the dark matter \cite{PBH1,PBH2,PBH3} has regained momentum \cite{kam,rep1,rep2} (see Ref. \cite{revPBH} for a recent review). While it is still unclear if PBHs can contribute to the totality of the dark matter \cite{c0,c1,c2}, there are various hints indicating that their abundance might be comparable to that of dark matter \cite{rev} and that they could provide the seeds for the cosmic structures \cite{carrsilk}.

While PBHs can be generated at the QCD epoch \cite{qcd} (for a recent analysis see Ref. \cite{bqcd}) or during phase transitions \cite{pt}, a popular mechanism for the formation of PBHs is within  the so-called ``spiky"  scenario in which PBHs are originated from the enhancement of the curvature power spectrum  below a certain length scale due to spiky features  \cite{s1,s2,s3}. Such an enhancement  can be produced either within single-field models of inflation, see for instance Refs. \cite{sone1,sone0,sone2} for some recent work, or through a spectator field which does  not contribute significantly to the total curvature perturbation   on large observable scales (which is ultimately responsible for the anisotropies in the cosmic microwave background anisotropies), but whose unique  role is precisely to enhance the power spectrum at small scales, see for example
Refs. \cite{stwo1,Carr,gauge,cs}. This spectator field can be even  the Higgs of the Standard Model \cite{ssm1,ssm2}.

Once the power spectrum of the curvature perturbation has been enhanced during inflation  from its $\sim 10^{-9}$ value at large scales to $\sim 10^{-2}$ on small scales, and after the fluctuations have been transferred to radiation during the reheating process after inflation, 
PBHs may form from sizeable  fluctuations  in the radiation density field if they  are able to overcome the resistance of the radiation pressure.

A perturbation of  fixed comoving size may not start  to collapse till it re-enters  the cosmological horizon. Therefore, the  size of a PBH at formation  is related to  the horizon length when the corresponding  perturbation enters the horizon and collapses.  Fluctuations  collapse immediately after horizon re-entry to form  PBHs  if they are sizeable enough. Indicating by  
$\zeta$  the gauge-invariant comoving curvature perturbation with power spectrum $P_\zeta$, one can define the smoothed version of it
\begin{eqnarray}
\label{sm}
\e_{R}(\vx)=\int \d^3 y\, W(|\vx-\vy|,R)\,  \e(\vy)
\end{eqnarray}
and  its variance  
\be
\sigma^2(R) =\int \frac{{\rm d}^3k}{(2\pi)^3}\,W^2(k,R)P_{\zeta}(k),
\ee
where the  smoothing  out is operated through    a  window function  $W(k,R)$  where one has to select the radius as  the comoving horizon length  $R_H=1/aH$.
A given region collapses to form a PBH if  this variance 
is  larger than a critical value $\zeta_c\simeq (0.05-1)$.  The actual value depends on the  equation of state $w$ of the fluid when the perturbation re-enters the horizon \cite{harada}
\be
\zeta_c=\frac{1}{3}\ln\frac{3(\chi_a-\sin\chi_a\cos\chi_a)}{2\sin^3\chi_a}\Big|_{\chi_a=\pi\sqrt{w}/(1+3w)}.
\ee
For radiation $w=1/3$  and  one obtains $\zeta_c\simeq 0.086$. Larger values are also used in the literature  \cite{bb,miller,hu}. A common value is   $\zeta_c\simeq 1.3$  and we will adopt it as representative from now on. 
Notice that one could use also a better  criterion \cite{bb} based on 
the density contrast   (during the  radiation era and on comoving slices) 
$\Delta(\vec x)=(4/9a^2H^2)\nabla^2\zeta(\vec x)$ for which the critical value is $\Delta_c\simeq 0.45$
 \cite{ls}\footnote{The use of the density contrast $\Delta(\vx)$ clarifies as well the use of the standardly adopted window function:  in the case of a spiky power-spectrum like the one we consider, most of the power is concentrated around the spikes; on the other hand, as we mentioned above, fluctuations cannot collapse as long as they are outside the horizon. Hence while the horizon acts as a sort of high-pass filter for $\Delta$, the combination of this filter with a spiky power-spectrum acts as a sort of band-pass filter which at each time selects the frequency range centred around the spike wavenumber which crosses the horizon at that time.}. At any rate, our formulae will be valid for both $\zeta(\vx)$ and $\Delta(\vx)$ and  their corresponding    power spectra on small scales must be rather large if a significant number of PBHs have to form.
The mass of a PBH at formation may be directly related to the number  $N$ of $e$-folds
before the end of inflation when  the  corresponding density fluctuation leaves the Hubble radius 
$M\simeq (m_{\rm P}^2/H){\rm exp}(2N)$ \cite{jn}, 
 where $m_{\rm P}$ is the Planck mass.

Under the assumption that the density contrast is a linear quantity obeying a Gaussian statistics, the primordial mass fraction $\beta_{\rm prim}(M)$ of the universe occupied by  PBHs formed at the time  $t_M$  is therefore given by 
\be
P(\zeta_{R_H}>\zeta_c)=\beta_{\rm prim} (M)=\int_{\zeta_c} \frac{{\rm d}\zeta_{R_H}}{\sqrt{2\pi}\,\sigma}e^{-\zeta_{R_H}^2/2\sigma^2},
\ee
where  too large fluctuations have to be disregarded as  the collapsing spacetime region behaves  as separate closed universe rather than a PBH \cite{PBH1}. 
The total contribution of  PBHs at radiation-matter equality is obtained by integrating the corresponding fraction $\beta(M,t_{\rm eq})=a(t_{\rm eq})/a(t_M)\beta_{\rm prim}(M,t_M)$ \cite{cg} 
at the time of equivalence 
\be
\Omega_{\rm PBH}(t_{\rm eq})=\int_{M_{\rm ev}(t_{\rm eq})}^{M(t_{\rm eq})}{\rm d}\ln M\, \beta(M,t_{\rm eq}), 
\ee
where $M_{\rm ev}(t_{\rm eq})\simeq 10^{-21} M_\odot$  is the lower mass  which has survived evaporation at equality
and $M(t_{\rm eq})$ is the horizon mass at equality. We warn the reader that, being them not the scope of this paper, in this expression various effects are not accounted for, such as the fact that the mass of the PBH is not precisely the mass contained in the corresponding horizon volume, but in reality obeys a  scaling relation with  initial perturbation \cite{jn},  that the  threshold is shape-dependent \cite{asp}, or finally that  the threshold amplitude and the final black hole mass depend on the initial density  profile of the perturbation \cite{misao}. Also, we will be dealing here only with the primordial PBH mass fraction.
From the time of equality to now, the PBH mass distribution will be altered by merging and accretion  \cite{c,cc}, processes which we have nothing to say on.

%
%
Now, defining  
\be
\nu(M)=\frac{\zeta_c}{\sigma(M)},
\ee
 the Gaussian mass fraction $\beta_{\rm prim}(M)$ for $\nu\gsim 5$ can be well approximated by 
 \be
 \label{omega}
\beta_{\rm prim}(M)\simeq \sqrt{\frac{1}{2\pi\nu^2}}
e^{-\nu^2/2}.
 \ee
 Since  PBHs are generated  through   very  large, but rare  fluctuations, their mass fraction at formation  is extremely sensitive to changes in the tail of the fluctuation distribution and therefore  to possible  non-Gaussianities in the density contrast fluctuations \cite{s3,byrnes,ngtwo,ng1,ng2,ng3,ng33,ng4,ng5,ng6}. This implies that  non-Gaussianities need to be accounted for within the spiky scenario as they can alter the initial mass fraction of PBHs in a dramatic way. 
 
 We are certainly not the first ones to make this point.  For instance, the   presence of a scale-invariant local non-Gaussianity, which is nowadays severely constrained by CMB anisotropy measurements \cite{planck},   can significantly alter the number density of PBHs  through   mode coupling  \cite{localNG,byrnes2}. In fact here we are rather  referring to that   non-Gaussianity which is almost inevitably generated at the same small wavelengths  where  the density perturbations are sizeable and  that in the literature is often   disregarded, most probably given the associated technical difficulties.
 
  To have the feeling of the level of non-Gaussianity
 associated to the large perturbations giving rise to the PBHs, let us consider the second-order gauge-invariant comoving curvature perturbation $\zeta_2$.
 During inflation, when the perturbations have already left the Hubble radius, it
 satisfies the equation \cite{second}
 \be
 \label{order}
 \dot{\zeta}_2=-\frac{H}{(\rho_0+P_0)}\left.\delta P_2\right|_{\rho}-\frac{2}{(\rho_0+P_0)}\left[\left.\delta P_1\right|_{\rho}-2(\rho_0+P_0)\zeta_1\right]\dot{\zeta}_1,
 \ee  
  where $\rho_0$ and $P_0$ are the background energy and pressure densities, $\left.\delta P\right|_{\rho}$ is the gauge-invariant non-adiabatic pressure perturbation (on uniform density hypersurfaces) and $\zeta_1$ is the linear gauge-invariant comoving curvature perturbation. In single-field models the non-adiabatic pressure perturbation
 vanishes and the curvature perturbation $\zeta$ is conserved at any order in perturbation theory on super-Hubble scales, consistently with Eq. (\ref{order}). Indeed, in single-field models the perturbations are generated at Hubble crossing and Eq. (\ref{order}) is not useful to estimate the level of non-Gaussianity associated to the fluctuations responsible for the PBHs and one has to resort to a different way of estimating the level of non-Gaussianity. However, in the case of spectator fields, the curvature perturbation is not conserved
 on super-Hubble scales since the non-adiabatic pressure perturbation is not zero. 
 In such a case,  Eq. (\ref{order}) signals that the same sizeable linear perturbations which will be responsible for the generation of
 the PBHs upon horizon re-entry inevitably give rise to sizeable  second-order
 perturbations and therefore to non-Gaussianity, that is $\zeta_2={\cal O}(1)\zeta_1^2$ barring cancellations. This simple argument shows that the level of non-Gaussianity
 in the large, but rare fluctuations associated to the PBHs nay be  not insignificant.

  The goal of this paper is to discuss the impact of non-Gaussianity on the initial PBH mass fraction. In particular, 
  
  \begin{enumerate} 
  
  \item we  follow  closely Ref. \cite{blm} and use a path-integral formulation to provide the probability that the density contrast is
  larger than a critical value within the horizon at formation time. The expressions are not original, but maybe known more in the community dealing with halo bias. They are exact and no approximations are taken;
  
  \item discuss the fine-tuning needed to achieve for the non-Gaussianity to play no role in the determination of the PBH mass function;
  
  \item compute in two popular classes of spiky models the impact of non-Gaussianity, showing that neglecting it leads to blunders in the estimation of the PBH mass fraction;
  
 \item discuss the impact of non-Gaussianity on the clustering of PBHs.

  \end{enumerate}
The paper is organised as follows. In section II we describe the technicalities based on the path-integral approach which allow to write exact expressions
for the various quantities of interest in the presence on non-Gaussianity. In section III we introduce the notion of fine-tuning associated to the
non-Gaussianity and in relation to the primordial PBH mass fraction. Sections IV and V contain two illustrative examples and section VI touches the notion of clustering and non-Gaussianity. Finally, conclusions are contained in section VII.

\section{The exact non-Gaussian PBH abundance}
\noindent
This section addresses the calculation of the exact non-Gaussian  PBH abundance through a path-integral approach. It has the virtue of not making any simplifying assumption or approximation, but at a price of being  rather technical. The uninterested reader can jump directly to the final result in Eq. (\ref{r1}) or its simplified version (\ref{final}).
We  follow
closely Ref. \cite{blm} which we reproduce the results of. In this sense this section does not contain any new finding, but  maybe results are derived in  a more transparent way. 

\vspace{-3mm}
\subsection{Basic definitions}
\noindent
Consider the overdensity field $\e(\vx)$ with a generic probability distribution $P[\e(\vx)]$. The partition function $Z[J]$ in the presence of an external source $J(\vx)$ is given by
\begin{eqnarray}
 Z[J]=\int [{D}\e(\vx)]P[\e(\vx)]e^{i\int \d^3 x J(\vx)\e(\vx)},
 \end{eqnarray} 
where the measure $[D\e(\vx)]$  is such that 
\begin{eqnarray}
\int [D\e(\vx)]P[\e(\vx)]=1. 
\end{eqnarray}
The functional Taylor expansion of the  partition function $Z[J]$  in powers of the source $J(\vx)$ specifies the correlators, while the corresponding expansion of $W[J]=\ln Z[J]$ generates the connected correlation functions, which we will denote by 
\be
\xi^{(N)}=\xi^{(N)}(\vx_1,\cdots,\vx_N).
\ee
 We have  already introduced the  smoothed   curvature perturbation  over a  scale $R$ in Eq. (\ref{sm}).  
In a similar manner, we define smoothed connected correlation function $\xi^{(N)}_{R}(\vx_1,\cdots,\vx_N)$ as 
\begin{eqnarray}
\xi^{(N)}_{R}(\vx_1,\cdots,\vx_N)=\int \xi^{(N)}_{R}(\vy_1,\cdots,\vy_n)\prod_{i=1}^N\d^3 y_iW(|\vx_i-\vy_i|,R),
\end{eqnarray}
so that 
\begin{eqnarray}
\xi^{(2)}_{R}(\vx,\vx)=\xi^{(2)}_{R}(0)=\sigma_{R}^2, \label{xx}
\end{eqnarray}
is the variance over the scale $R$. For later use, we also define the normalised correlators 
\begin{eqnarray}
w^{(N)}_{R}(\vx_1,\cdots,\vx_N)=\sigma_{R}^{-N}\, \xi^{(N)}_{R}(\vx_1,\cdots,\vx_N)
\end{eqnarray}
where in particular
\begin{eqnarray}
w^{(2)}_{R}(0)=1.
\end{eqnarray}
Let us also recall that the functional  $W[J]$ is written in terms of the connected correlation functions $\xi^{(N)}$ as 
\begin{eqnarray}
W[J]=\sum_{n=2}^\infty\frac{(-1)^n}{n!}
\int \d^3 x_1\cdots \int \d^3 x_N \, \xi^{(N)}(\vx_1,\cdots,\vx_N)J(\vx_1)\cdots J(\vx_N).
\label{wj}
\end{eqnarray}

\subsection{Threshold statistics}
\noindent
We now wish to characterise the   threshold statistics by requiring   that $\e_{R}$ exceeds  a certain threshold $\zeta_c$, which we can write as   $\e_{R}>\nu\sigma_{R}$. Then the peak overdensity $\rho_{\nu,R}(\vx)$ is defined as 
\begin{eqnarray}
 \rho_{\nu,R}(\vx)=\Theta(\e_{R}(\vx)-\nu\sigma_{R}),
 \end{eqnarray} 
where $\Theta(x)$ is the Heaviside step-function. 
The joint probability that in all points, $(\vx_1,\cdots, \vx_N)$, the overdensity is above  the threshold is given by
\begin{eqnarray}
\Pi_{\nu,R}^{(N)}(\vx_1,\cdots,\vx_N)=\bigg\langle\prod_{i=1}^N\rho_{\nu,R}(\vx_i)\bigg\rangle=\int [D\e(\vx)]P[\e(\vx)]
\prod_{i=1}^N\Theta\Big(\e_{R}(\vx_i)-\nu\sigma_{R}\Big).
\end{eqnarray}
Since $\d\Theta(\vx)/\d x=\delta(\vx)$ and $2\pi  \delta(\vx)=\int_{-\infty}^\infty\d \phi\, e^{i\phi x}$,  the $\Theta$-function may be expressed as 
\begin{eqnarray}
\Theta(\vx)=\int_{-x}^\infty \d a \int_{-\infty}^\infty\frac{\d\phi}{2\pi} \, e^{i\phi a},
\end{eqnarray}
and therefore, 
\begin{align}
\Pi_{\nu,R}^{(N)}&=\int [D\e(\vx)]P[\e(\vx)]\prod_{i=1}^N\int_{\nu \sigma_{R}}^\infty \d a_i\int_{-\infty}^\infty
\int_{-\infty}^\infty  \frac{\d\phi_i}{2\pi}
\,e^{i\phi_i(\e_{R}(\vx_i)-a_i)}\nonumber \\
&= (2\pi)^{-N}\int_{\nu \sigma_{R}}^\infty \d a_1\cdots \int_{\nu \sigma_{R}}^\infty \d a_N
\int_{-\infty}^\infty\d \phi_1\cdots \int_{-\infty}^\infty\d \phi_N \, \exp{\Big[-i\sum_{i=1}^N\phi_ia_i\Big]}Z[J]\nonumber \\
&=(2\pi)^{-N}\sigma_{R}^N\int_{\nu }^\infty \d a_1\cdots \int_{\nu}^\infty \d a_N
\int_{-\infty}^\infty\d \phi_1\cdots \int_{-\infty}^\infty\d \phi_N \, \exp{\left(-i\sigma_{R}\sum_{i=1}^N\phi_ia_i\right)}Z[J], \label{al}
\end{align}
where $J$ is defined as $J(\vx)=\sum_{i=1}^N J_i$ with $J_i(\vx_i,\vx)=\phi_i W(|\vx-\vx_i|,R)$, and   in the third line in Eq. (\ref{al}) we have redefined $a_i\to a_i/\sigma_{R}$. Using  Eq. (\ref{wj}) we may  express $\ln Z[J]$ as 
\begin{equation}
\begin{aligned}
\ln Z[J]&=& \sum_{n=2}^\infty\frac{(-1)^n}{n!}\int \d^3 \vy_1\cdots \int \d^3 \vy_n \,
\sum_{i_1=1}^N\cdots \sum_{i_n=1}^N J_{i_1}(\vy_1,\vx_1)\cdots J_{i_n}(\vy_n,\vx_n)
 \xi^{(n)}(\vy_1,\cdots,\vy_n) \\
 &=& \sum_{n=2}^\infty\frac{(-1)^n}{n!}
\sum_{i_1=1}^N\cdots \sum_{i_n=1}^N \phi_{i_1}\cdots \phi_{i_n}
 \xi^{(n)}_{R}(\vx_{i_1},\cdots,\vx_{i_n})
 =\sum_{n=2}^\infty \frac{(-1)^n}{n!}\sum_{i_1,\cdots,i_n=1}^N 
 \xi^{(n)}_{R,[i_n]}\prod_{k=1}^n\phi_{i_k},
\end{aligned}
	\end{equation}
 where 
 \begin{eqnarray}
 \xi^{(n)}_{{R},[i_n]}=\xi^{(n)}_{{R}}(\vx_{i_1},\cdots,\vx_{i_n}).
 \end{eqnarray}
We may put the last expression in a  more explicit form using Eq. (\ref{xx})
 \begin{eqnarray}
 \ln Z[J]&=&\frac{1}{2}\phi_1^2 \xi^{(2)}_{R}(\vx_1,\vx_1)+\frac{1}{2}\phi_2^2 \xi^{(2)}_{R}(\vx_{2},\vx_{2})+\cdots\nonumber\\ &&
 +\frac{1}{2} \phi_1\phi_2 \xi^{(2)}_{R}(\vx_1,\vx_{2})-\frac{1}{3} \phi_1\phi_2 \phi_3 \xi^{(3)}_{R}(\vx_1,\vx_{2},\vx_{3})+\cdots\nonumber \\
 &=&
\frac{1}{2} \sigma_{R}^2\sum_{i=1}^N\phi_i^2+\frac{1}{2}\sum_{i_1\neq i_2}^N
\phi_{i_1}\phi_{i_2}\xi^{(2)}_{{R},[i_2]}-\frac{1}{6}
\sum_{i_1,i_2,i_3=1}^N\phi_{i_1}\phi_{i_2}\phi_{i_3}\xi^{(3)}_{{R},[i_3]}
+\cdots\, .
\end{eqnarray}
Using the above expression and since 
\begin{eqnarray}
\phi_i e^{i \sum_{i=1}^N \phi_i a_i}=-i \frac{\partial}{\partial a_i} e^{i \sum_{i=1}^N \phi_i a_i},
\end{eqnarray}
we find 
\begin{equation}
\begin{aligned}
\Pi_{\nu,R}^{(N)}=&(2\pi)^{-N}\sigma_{R}^N\int_{\nu }^\infty \d a_1\cdots \int_{\nu}^\infty \d a_N
\int_{-\infty}^\infty\d \phi_1\cdots \int_{-\infty}^\infty\d \phi_N \, \times 
\\
& \times \,\,
\exp\left\{\sum_{n=2}^\infty \frac{(-1)^n}{n!}
\sum_{i_1,\cdots,i_n=1}^N \xi^{(n)}_{{R},[i_n]}\prod_{i=1}^n
\frac{\partial}{\partial a_i}\right\}
\exp{\left(-\frac{1}{2}\sigma_{R}^2\sum_{i=1}^N\phi_i^2-i\sigma_{R}\sum_{i=1}^N\phi_ia_i\right)}, 
\end{aligned}
	\end{equation}
which upon   Gaussian integration over $\phi_i$'s,  becomes
\begin{equation}
\begin{aligned}
\Pi_{\nu,R}^{(N)}=&(2\pi)^{-N/2}\int_{\nu }^\infty \d a_1\cdots \int_{\nu}^\infty \d a_N
\, \times  \label{pp}
\\
&\times \,\,
\exp\left\{\sum_{n=2}^\infty \frac{(-1)^n}{n!}\sum_{[i_n]=1}^N 
w^{(n)}_{{R},[i_n]}\prod_{i=1}^n
\frac{\partial}{\partial a_i}\right\}
\exp{\left(-\frac{1}{2}\sum_{i=1}^N a_i^2\right)}. 
\end{aligned}
	\end{equation}
To calculate the expression (\ref{pp}), we have used the formula
\begin{eqnarray}
\exp\left(\sum_{n=2}^\infty \frac{1}{n!}S[n]\right)&=&1+\frac{1}{2!}S[2]+\frac{1}{3!}S[3]+\frac{1}{4!}\Big(3S[2]^2+S[4]\Big)+\frac{1}{5!}\Big(10S[2]S[3]+S[5]\Big)\cdots\nonumber \\
&=&
\sum_{\ell=0}^\infty\sum_{\hat{p}[\ell]}
\prod_{\substack{p_1m_1+\cdots m_rp_r=\ell\\p_i,m_i\geq 0}}
\frac{S[m_1]^{p_1}\cdots S[m_r]^{p_r}}{m_1!\cdots m_r! p_1!\cdots p_r!},
\label{for}
\end{eqnarray}
where $\hat{p}[\ell]$ denotes the partitions of the integer $\ell$ into number $\ell_i$ with $\ell_i>1$. Therefore, each $\ell$-th term in the sum contains $p(\ell)-p(\ell-1)$ terms, where $p(\ell)$ is the number of partitions of $\ell$  into numbers $\ell_i$ but with 
$\ell_i\geq 1$ now.  

By defining 
\begin{eqnarray}
\Xi(\vx_{i_1},\cdots, \vx_{i_\ell})= \sigma_{R}^{-\ell}\, \sum_{\hat{p}[\ell]}\prod_{\substack{p_1m_1+\cdots m_rp_r=\ell\\p_i,m_i\geq 0}}
\frac{\ell!\,{\xi_R^{(m_1)}}^{p_1}\cdots {\xi_R^{(m_r)}}^{p_r}}{m_1!\cdots m_r! p_1!\cdots p_r!}, \label{XXX}
\end{eqnarray}
 we may express (\ref{pp}) as 
\begin{eqnarray}
\Pi_{\nu,R}^{(N)}=(2\pi)^{-N}\sigma_{R}^N\int_{\nu }^\infty \d a_1\cdots \int_{\nu}^\infty \d a_N
\left\{\sum_{\ell=0}^\infty\frac{1}{\ell!}\,  \Xi(\vx_{i_1},\cdots, \vx_{i_\ell})\partial_{i_1\cdots i_\ell}\exp{\left(-\frac{1}{2}\sum_{i=1}^N a_i^2\right)}\right\}, \label{ppp}
\end{eqnarray}
where $\partial_{i_1\cdots i_\ell}=\partial/\partial a^{i_1}\cdots\partial/\partial a^{i_\ell}$.
Le us now denote by $P_{n,N}$ the set of all partitions of the integer $n$ into  at most $N$ parts and let 
$[n]_N=\{n_1,\cdots,n_N\}$ be its elements.
Then, if we denote 
\begin{eqnarray}
h_m(\nu)=\frac{1}{\sqrt{2^m\pi}}e^{-\nu^2/2}H_{m-1}\left(\frac{\nu}{\sqrt{2}}\right),~~~~h_0(\nu)=\frac{1}{2}
{\rm Erfc}\left(\frac{\nu}{\sqrt{2}}\right),
\end{eqnarray}
where $H_{m-1}$ are Hermite polynomials and ${\rm Erfc}(x)=\sqrt{2}/\pi\int_x^\infty \exp(-t^2)\d t$ is the complementary error function, we find that 
Eq. (\ref{ppp}) 
can be re-expressed as 

\begin{eqnarray}
\Pi_{\nu,R}^{(N)}(\vx_1,\cdots,\vx_N)= \sum_{\ell=0}^\infty\frac{1}{\ell!}\left\{\sum_{\{\ell_1,\cdots,\ell_N\}\in P_{\ell,N}} \frac{\ell!}{n_1!\cdots n_N!}
\Xi^{\{\ell_1,\cdots,\ell_N\}}
\prod_{\ell_r}^N h_{\ell_r}(\nu)\right\},
\end{eqnarray}
where $\Xi^{\{\ell_1,\cdots,\ell_N\}}$ denotes 
the sum of $\Xi(\vx_{i_1},\cdots,\vx_{i_\ell})$'s at $\ell_1, \cdots \ell_N$ points. For $N=3$ and $\ell=4$, the possible partitions are $(4,0,0),(3,1,0),(2,2,0),(2,1,1)$ and then we have for example,
\begin{eqnarray}
&&\Xi^{\{4,0,0\}}=\Xi(\vx_1,\vx_1,\vx_1,\vx_1)+\Xi(\vx_{2},\vx_{2},\vx_{2},\vx_{2})+\Xi(\vx_{3},\vx_{3},\vx_{3},\vx_{3}),\\
&&\Xi^{\{2,2,0\}}=\Xi(\vx_1,\vx_1,\vx_{2},\vx_{2})+\Xi(\vx_1,\vx_1,\vx_{3},\vx_{3})+\Xi(\vx_{2},\vx_{2},\vx_{3},\vx_{3})
\end{eqnarray}
and 
\begin{eqnarray}
\Xi^{\{2,1,1\}}=\Xi(\vx_1,\vx_1,\vx_{2},\vx_{3})+\Xi(\vx_{2},\vx_{2},\vx_1,\vx_{3})+\Xi(\vx_{3},\vx_{3},\vx_1,\vx_{2}).
\end{eqnarray}
\subsection{The non-Gaussian one-point function or probability to be above the threshold}
\noindent
We are finally ready to collect the results to find the probability for a single density contrast to be above the threshold. It can be directly calculated by evaluating the one-point statistics
\begin{equation}
\begin{aligned}
\label{finalc}
P(\zeta_R>\zeta_c)=\langle\rho_{\nu,R}(\vx)\rangle=&\Big< \Theta(\zeta_{R}(\vx)-\nu\sigma_{R})\Big> \\
 =&
(2\pi)^{-1/2}\int_{\nu }^\infty \d a
\,
\exp\left\{\sum_{n=3}^\infty \frac{(-1)^n}{n!} w^{(n)}_{{R}}(0)
\frac{\partial^n}{\partial a^n}\right\}
\exp{\left(-\frac{1}{2}a^2\right)}\\
=&
(2\pi)^{-1/2}\int_{\nu }^\infty \d a
\,\left(1-\frac{1}{3!}w^{(3)}(0)\frac{\d^3}{\d a^3}+
\frac{1}{4!}w^{(4)}(0)\frac{\d^4}{\d a^4}+\cdots\right)\exp{\left(-\frac{1}{2}a^2\right)}\\
=& h_0(\nu)+\frac{1}{\sqrt{4\pi}\,3!}w^{(3)}(0)e^{-\nu^2/2}H_{2}\left(\frac{\nu}{\sqrt{2}}\right)
+\frac{1}{\sqrt{16\pi}\,4!}w^{(4)}(0)e^{-\nu^2/2}H_{3}\left(\frac{\nu}{\sqrt{2}}\right)+\cdots\, ,
 \\
\end{aligned}
	\end{equation}
or
\be
\boxed{
P(\zeta_R>\zeta_c)= h_0(\nu)+\frac{e^{-\nu^2/2}}{\sqrt{2\pi}}\sum_{n=3}^\infty 
\frac{1}{2^{\frac{n}{2}}n!}\Xi_n(0)H_{n-1}\left(\frac{\nu}{\sqrt{2}}\right), }\label{r1}
\ee
where the argument $(0)$ means that all the correlation functions are computed at the same point, 
\begin{eqnarray}
\Xi_n(0)=\Xi(\vx,\vx,\cdots, \vx),
\end{eqnarray}
and $\Xi(\vx_1,\vx_{2},\cdots, \vx_n)$ is given by Eq. (\ref{XXX}) with $\xi^{(2)}_R$ omitted. For instance, we have
\begin{eqnarray}
&&\Xi_3(0)=w^{(3)}(0), ~~~~\Xi_4(0)=w^{(4)}(0), ~~~~\Xi_5(0)=w^{(5)}(0),
\nonumber \\
&&\Xi_6(0)=w^{(6)}(0)+10 w^{(3)}(0)^2, ~~~\Xi_7(0)=w^{(7)}(0)+35 w^{(3)}(0)
w^{(4)}(0).
\end{eqnarray}
We draw the attention of the reader to the fact that the expression (\ref{r1}) is exact and no approximations have been taken along the way.
Nevertheless, it simplifies in the case of large threshold $\nu\gg1$.  In this limit, using the asymptotic behaviour ($\nu\gg 1$) of the Hermite polynomials and the complementary error function 
\begin{eqnarray}
&&H_n\left(\frac{\nu}{\sqrt{2}}\right)= 2^{\frac{n}{2}}\nu^n \Big(1 +{\cal O}(\nu^{-2})\Big)\nonumber \\
&&h_0(\nu)=\frac{1}{2}{\rm Erfc}\left(\frac{\nu}{\sqrt{2}}\right)= 
\frac{e^{-\nu^2/2}}{\sqrt{2\pi\nu^2}}\Big(1 +{\cal O}(\nu^{-2})\Big), 
\end{eqnarray}
we find that Eq. (\ref{r1}) reduces to 
\begin{equation}
\begin{aligned}
\boxed{
P(\zeta_R>\zeta_c)=\frac{e^{-\nu^2/2}}{\sqrt{2\pi \nu^2}}\exp\left\{\sum_{n=3}^\infty
\frac{(-1)^{n}}{n!}w^{(n)}_{R}(0) \, \nu^n\right\}=\frac{1}{\sqrt{2\pi \nu^2}}\exp\left\{-\nu^2/2+\sum_{n=3}^\infty
\frac{(-1)^{n}}{n!}\xi^{(n)}_{R}(0) \, (\zeta_c/\sigma_{R}^2)^n\right\},}
\label{final}
\end{aligned}
	\end{equation}
where the argument $(0)$ means that all the correlation functions are computed at the same point. If the smoothing radius is taken to be 
$R_H$, this expression becomes the mass fraction $\beta_{\rm prim}(M)$. 
Apart from the limit $\nu\gg 1$, it is exact as it  contains all the non-Gaussian correlators for which no assumption needs to be taken. If one wishes to relax even the assumption of large $\nu$, one can resort to Eq. (\ref{r1}).

\section{The impact of non-Gaussianity on the PBH abundance}
\noindent
After this technical detour, we are ready to ask under which circumstances  non-Gaussianity alters the predictions of the primordial abundance of black holes in a significant way. To answer this question one of course should   look at her/his own preferred model and estimate the amount of non-Gaussianity
through the non-vanishing  correlators with order larger than two. However, one can take inspiration by what is routinely done in  high energy physics and introduce the notion of fine-tuning, that is how much sensitive is a given observable on the parameters it depends upon. For instance, in supersymmetric models, where the phenomenon of electroweak symmetry breaking is induced radiatively by the running with energy, one may compute the response of an observable such as the 
mass squared of the $Z$-boson $m_Z^2$ to a change of a parameter of the model defined at some high energy \cite{bg}. 

In full similarity, we  can ask ourselves how sensitive is $\beta_{\rm prim}(M)$ to the various cumulants, defined by the relations
\be
S_n=\frac{\xi^{(n)}_{\nu,R_H}(0)}{\left(\xi^{(2)}_{\nu,R_H}(0)\right)^{n-1}}=\frac{\overset{n-{\rm times}}{\langle\overbrace{\zeta_{R_H}(\vx)\cdots\zeta_{R_H}(\vx)}\rangle} }{\sigma^{2(n-1)}_{R_H}}.
\ee
We may therefore define the fine-tuning $\Delta_n$  to be the response of the PBH abundance to the 
introduction of the $n$-th cumulant
\be
\Delta_n=\frac{\d\ln \beta_{\rm prim}(M)}{\d\ln S_n}.
\ee
Given the particular structure of the non-Gaussian correction to the PBH abundance, one can immediately understand the physical significance of this
fine-tuning parameter. Indeed, in the presence of the $n$-th cumulant, one can express the new non-Gaussian PBH abundance (\ref{final}) in terms 
of    the Gaussian abundance 
\begin{eqnarray}
 \frac{\beta^{\rm NG}_{\rm prim} (M)}{\beta^{\rm G}_{\rm prim}(M)}=e^{\Delta_n}.
 \end{eqnarray}
 This implies that the PBH abundance is exponentially sensitive to the non-Gaussianity unless $\Delta_n$ is  in absolute value smaller than unity
 \be
|\Delta_n|\lsim 1.
\ee
Inspecting  Eqs. (\ref{omega}) and (\ref{final}), we see that
\be
| \Delta_n|=\frac{1}{n!}\left(\frac{\zeta_c}{\sigma_{R_H}}\right)^2|S_n|\zeta_c^{n-2}.
\ee
This tells us that non-Gaussianity alters exponentially the Gaussian prediction for the PBH abundance unless
\be
\label{condition}
\boxed{
\left|S_n\right|\lsim \left(\frac{\sigma_{R_H}}{\zeta_c}\right)^2\frac{n!}{\zeta_c^{n-2}}.}
\ee
To investigate how restrictive this condition is, we will analyse two representative classes of spiky models, one based on a single-field and the other based on the presence of a spectator field.

\section{PBHs from single-field models of inflation}
\noindent
In single-field models of inflation the power spectrum of the comoving curvature perturbation is given by \cite{lrreview}
\be
P_\zeta(k)=\frac{H^2}{8\pi^2m_{\rm P}^2\epsilon}\frac{1}{k^3},
\ee
where $\epsilon=-\d\ln H/\d N$ is one of the slow-roll parameters and $N$ is the number of $e$-folds till the end of inflation. Since the generation of PBHs requires the jumping of the value of the power spectrum of about seven orders of magnitude from its value on CMB scales,  without even specifying the single-field model of inflation,  one may conclude that there must be a  violation of the slow-roll condition $|\Delta \ln\epsilon/\Delta N|\gsim {\cal O}(1)$  \cite{hu}. In particular, this happens when the inflaton field goes through  an inflection point in the scalar potential, see Fig. 1, thus   producing a sizeable resonance in the power spectrum of the curvature perturbation.
 Correspondingly, slow-roll conditions are broken and the spike in the  curvature perturbation may   lead to a copious production of PBHs, see for instance Refs. \cite{sone1,sone2} for some recent proposals.
 
When the inflaton experiences a plateau in its potential, since $\epsilon$ must be extremely small, a short period of Ultra Slow-Roll  (USR) is achieved during which  the equation of  motion
of the inflaton background $\phi_0$ reduces to \cite{u1,u2,u3,u4,u5,u6}
\be
\ddot{\phi}_0+3 H\dot{\phi}_0\simeq 0,
\ee
and the piece $\d V(\phi_0)/\d  \phi_0$ is negligible so that there is an approximate shift symmetry in the inflaton field. The slow-roll parameter 
\be
\eta=\frac{\dot\epsilon}{\epsilon H}=2\frac{\ddot{\phi}_0}{\dot{\phi}_0 H}+2\epsilon\simeq -6
\ee
is therefore sizeable, signalling the breaking of the slow-roll conditions.
During such a phase of USR inflation, 
\noindent
\begin{figure}[h!]
    \begin{center}
      \includegraphics[scale=.5,angle=360]{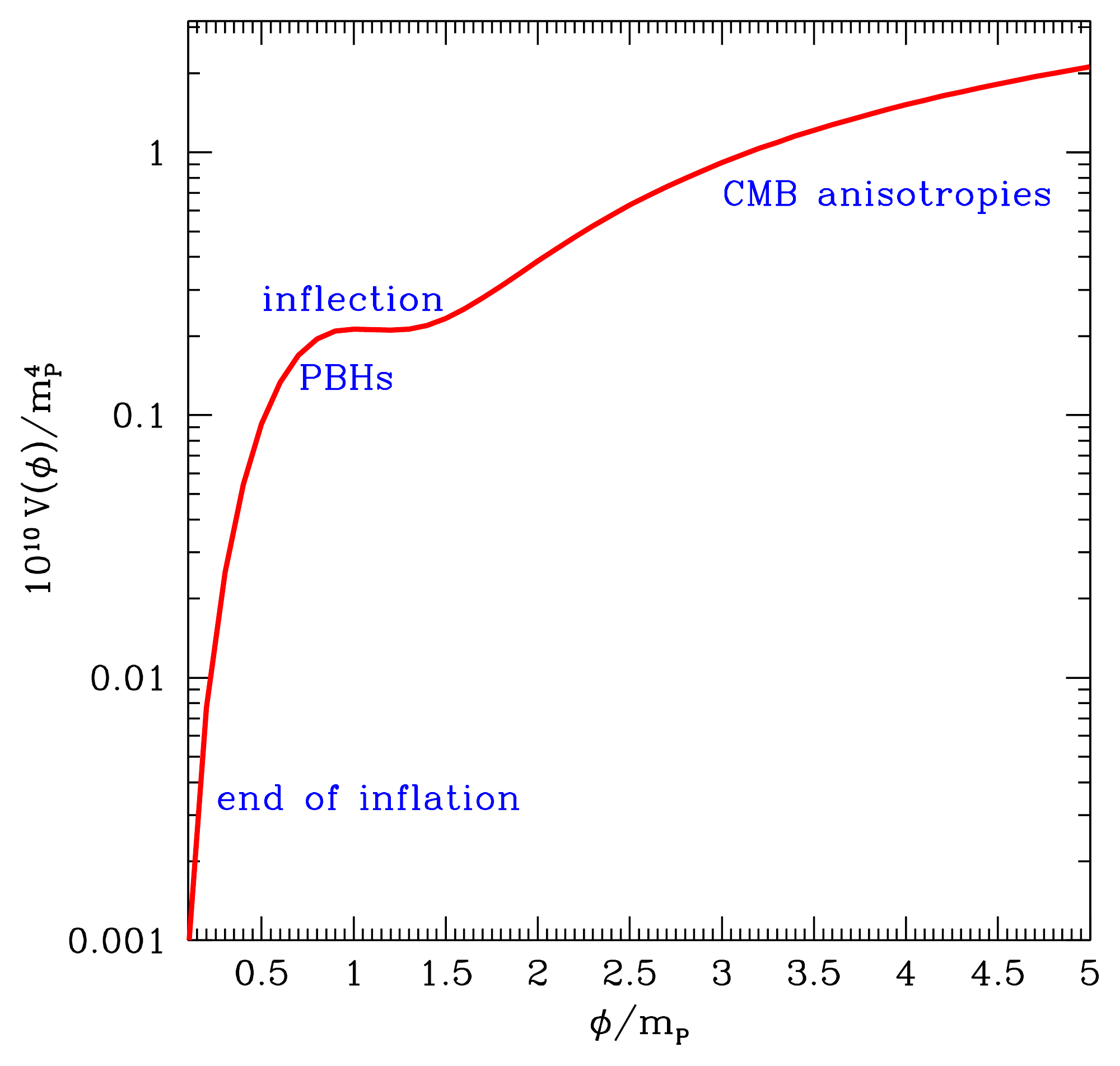}
    \end{center}
     \caption{\small A representative inflaton potential with an inflection point where the sizeable perturbations leading to the PBHs can be generated.}  \label{fig1}
\end{figure}
\noindent 
\be
\epsilon=\frac{3}{2}\frac{\dot{\phi}_0^2}{V(\phi_0)}\sim a^{-6},
\ee
and the comoving curvature perturbation  increases as 
\be
 P_\zeta\sim e^{6\Delta N},
\ee
where $\Delta N$ parametrises the duration of the USR phase. Thanks to the approximate shift symmetry as well as the dilation symmetry, one can characterise the bispectrum of the perturbations generated during the USR  phase. The non-Gaussianity turns out to be  completely local ($\delta\phi$  is  basically a massless scalar field on the de Sitter background in the limit $\epsilon\lll 1$ and  non-Gaussianity is generated after Hubble radius crossing) \cite{u3,u4,pajer}
\be
\Big<\zeta_{\vec{k}_1} \zeta_{\vec{k}_2} \zeta_{\vec{k}_3}\Big>=3\cdot(2\pi)^3\delta^{(3)}(\vec{k}_1+\vec{k}_2+\vec{k}_3) \left[P_\zeta(k_1)P_\zeta(k_2)+{\rm two}\,\,{\rm terms}\right].
\ee
The corresponding skewness reads 
\begin{eqnarray}
\label{nine}
S_3&=&\frac{1}{\sigma_{R_H}^4}\int\frac{\d^3 k_1}{(2\pi)^3}\int\frac{\d^3 k_2}{(2\pi)^3}\int\frac{\d^3 k_3}{(2\pi)^3}W(k_1,R_H)W(k_2,R_H)W(k_3,R_H)\Big<\zeta_{\vec{k}_1} \zeta_{\vec{k}_2} \zeta_{\vec{k}_3}\Big>
\simeq 9,
\end{eqnarray}
where we have  neglected the effect of the window functions in the momentum integrals\footnote{Had we used  a volume-normalised Gaussian window function $W(k,R_H)={\rm exp}(-k^2 R_H^2/2)$ and a spiky power spectrum of the form $P_\zeta(k)\simeq (A^2/k^3)\delta(\ln k/k_*)$ peaked around
the wavenumber $k_*$ and with amplitude $A^2$, we would have obtained $S_3\simeq 9\cdot{\rm sinh}(k_*^2R_H^2)/(k_* R_H)^2$, which differs
at $R_H=1/k_*$ from the result (\ref{nine}) only by a factor 1.17.}.
%
%
%
%
%
Since the condition (\ref{condition}) for non-Gaussianity to play no significant  role is
\be
S_3\lsim 0.06 \left(\frac{\sigma_{R_H}^2}{0.02}\right) \left(\frac{1.3}{\zeta_c}\right)^3,
\ee
where we have normalised $\sigma_{R_H}^2$ to the value necessary to get the right abundance of dark matter from PBHs \cite{hu} (neglecting accretion), we see that the Gaussian estimate on the PBH mass fraction is hardly trustable.  Of course this result strongly depends on the values of $\sigma_{R_H}^2$ and $\zeta_c$, but it signals a potential danger. 

In fact, one can do better at this stage. We can use the $\delta N$ formalism \cite{rep3,rep4,deltaN} to evaluate
the higher-order cumulants. In the $\delta N$ formalism  the scalar field fluctuations are quantized on the flat slices and related to the comoving curvature perturbation $\zeta$ through the relation $\zeta=\delta N$. During the USR phase and in the case of a sharp transition back to the slow-roll phase, one obtains \cite{u3}
\be
N(\phi_0,\dot{\phi}_0)=\frac{1}{3}\ln\left[\frac{\dot{\phi}_0}{\dot{\phi}_0+3H(\phi_0-\phi_e)}\right],
\ee
where $\phi_e$ is the value of the field at the end of the USR phase. Notice in particular that one has to retain the dependence on
$\dot{\phi}_0$ since slow-roll is badly violated. Since $\delta\phi$ is constant to good accuracy, one can neglect $\delta\dot\phi$ and obtain
\begin{eqnarray}
\zeta(\vx)=\delta N(\vx)&=&N\left(\phi_0+\delta\phi(\vx),\dot{\phi}_0+\delta\dot{\phi}(\vx)\right)-N\left(\phi_0,\dot{\phi}_0\right)\nonumber\\
&=&\sum_{n\geq 1}\frac{1}{n!}\frac{\partial^n N}{\partial \phi_0^n}(\delta\phi(\vx))^n=\sum_{n\geq 1}(-1)^{n}\frac{3^{n-1}}{n}\left(\frac{H}{\dot{\phi}_0+3H(\phi_0-\phi_e)}\right)^n(\delta\phi(\vx))^n,
\end{eqnarray}
or
\begin{eqnarray}
\zeta(\vx)&=&\sum_{n\geq 1}{\cal N}_n(\zeta_1(\vx))^n,\nonumber\\
{\cal N}_n &=&\frac{3^{n-1}}{n},
\end{eqnarray}
where $\zeta_1(\vx)$ is the linear curvature perturbation. If one considers the contribution to the cumulant $S_n$ coming from the connected piece of the form
\begin{equation}
\begin{aligned}
S_n\supset s_n=\frac{n{\cal N}_{n-1}}{\sigma_{R_H}^{2(n-1)}}\int\frac{\d^3 k_1}{(2\pi)^3}W(k_1,R_H)\cdots\int\frac{\d^3 k_n}{(2\pi)^3}W(k_n,R_H)\Big<\zeta_{1,\vec{k}_1} \cdots\zeta_{1,\vec{k}_{n-1}} (\zeta_1)^{n-1}_{\vec{k}_n}\Big>\simeq n!\,{\cal N}_{n-1},
\end{aligned}
	\end{equation}
where again  we have  neglected the effect of the window functions in the momentum integrals, 
it  turns out that one can resum the series in the expression (\ref{final}) as long as $\zeta_c\leq 1/3$
\begin{eqnarray}
\sum_{n=3}^\infty
\frac{(-1)^{n}}{n!}\xi^{(n)}_{R}(0) \, (\zeta_c/\sigma_{R}^2)^n&\supset&\frac{1}{\sigma_{R_H}^2} \sum_{n=3}^\infty
\frac{(-1)^{n}}{n!}s_n \, \zeta_c^n\nonumber\\
&=&\frac{1}{\sigma_{R_H}^2}\sum_{n=3}^\infty
(-1)^{n} \frac{3^{n-2}}{n-1}\, \zeta_c^n\nonumber\\
&=&-\frac{\zeta_c^2}{\sigma_{R_H}^2}\left[1-\frac{1}{3\zeta_c}\ln\left(1+3\zeta_c\right)\right].
\end{eqnarray}
This contribution shifts the Gaussian exponential in the mass fraction in the expression (\ref{final}) from (in this expression one should
intend the variance $\sigma^2_{R_H}$ as the renormalized one once the non-Gaussian contributions are as well summed up \cite{byrnes,diagrammatic})
\be
\label{hi}
P_1(\zeta_{R}>\zeta_c)=\frac{e^{-\nu^2/2}}{\sqrt{2\pi \nu^2}}\xrightarrow{\zeta_c= 1/3} P_{\rm NG}(\zeta_{R}>\zeta_c)\simeq\frac{e^{-1.6\,\nu^2/2
}}{\sqrt{2\pi \nu^2}}
\ee
This is not at all an insignificant effect for $\nu\gg 1$. For instance one can ask what is  the impact on the PBH abundance at formation. We follow
 Ref. \cite{hu}  and we set $\beta_{\rm prim}(M)$ such that it corresponds today to a totality of the dark matter in the form of PBHs.  Choosing
 $\zeta_c\simeq 1/3$, a Gaussian probability would give $\sigma_{R_H}\gsim 0.036$, while the non-Gaussian  result (\ref{hi}) would deliver
a result smaller than about ten orders of magnitude. This shows the large sensitivity of the non-Gaussian result to the cosmological observables.
 Of course, this is not the final result as there are other
contributions to the cumulants we have not accounted for. For instance, $S_4$ receives a connected contribution from the correlator $\langle\zeta_1(\vx)\zeta_1(\vx)\zeta^2_1(\vx)\zeta^2_1(\vx)\rangle$. We have not been able to resum  all the pieces. Despite their incompleteness and the fact that  the details have to be worked out model by model,  our findings indicate
that non-Gaussianity plays a fundamental role in determining the final PBH abundance.

Notice that we have been  working under the assumption of a sudden transition between the USR phase and the slow-roll subsequent stage. It has been recently noticed that the non-Gaussianity is sensitive to the details of the transition from USR phase to the standard slow-roll phase \cite{rev1}. The skewness takes the generic  value $S_3\simeq 9(4(\eta_V-3)\eta_V+h^2+4\eta_V h)/(2\eta_V+h-6)^2$, where $h=6\sqrt{\epsilon_V}/\pi_e$, $\pi=\d\phi_0/\d N$ and  $\epsilon_V$ and $\eta_V=V''/3H^2$  are  evaluated at the slow-roll phase. For $|h|\gg 1$ one recovers the sharp transition case, while for a smooth transition or a sharp transition,  but small $h$, one finds the value 
  $S_3\simeq\eta_V$. Said in other words, our results show that a smooth transition
is preferable in order to have a trustable  Gaussian estimate.

\section{PBHs from the axion-curvaton model}
\noindent
The second example we wish to illustrate falls in the category of spiky models where the comoving curvature power spectrum is enhanced at small scales
through a spectator field \cite{stwo1}. The idea is that there is a curvaton-like field $\sigma$ \cite{curvaton} with quadratic potential 
\be
\label{qq}
V(\sigma)=\frac{1}{2}m^2\sigma^2,
\ee
such that the total curvature perturbation during inflation 
reads  (in the flat gauge)
\be
\zeta=H\frac{\delta\rho}{\dot \rho}=
\frac{\dot{\rho}_{\phi}}{\dot\rho}\zeta_\phi+\frac{\dot{\rho}_\sigma}{\dot\rho}\zeta_\sigma,\,\,\,\,\,\,\,\,\,\,\zeta_{\phi,\sigma}=H\frac{\delta\rho_{\phi,\sigma}}{\dot\rho_{\phi,\sigma}},
\ee
where  $\zeta_\sigma$ is the curvaton perturbation and 
 $\zeta_\phi$   is responsible for the anisotropies in the CMB anisotropies on large scales. The two contributions are assumed to be comparable
 at a scale $k_c\simeq$ Mpc$^{-1}$.
The curvaton is in fact an axion-like field. It is  related to the phase $\theta$ of a complex field whose modulus is the inflaton field $\phi_0$ with potential
\be
V(\phi_0)=\frac{1}{2}\lambda H^2\phi_0^2,
\ee
with $1<\lambda\leq9/4$. The inflaton rolls relatively fast towards its minimum $\phi_*$ which is reached when the wavenumber $k_*$ is exiting the Hubble radius and the curvaton field becomes well-defined, $\sigma(\vx)=\phi_*\theta(\vx)$. For scales $k<k_*$, when the inflaton is still rolling down its potential,
the fluctuations of the curvaton inherit a strong scale-dependence, $\delta\theta_{\vk}\simeq (H/2\pi\phi_0(t_k))$, where $t_k$ is the time at which the wavelength $1/k$ exits the Hubble radius. The corresponding spectral index is given by
\be
n_\sigma-1\simeq 3-3\sqrt{1-\frac{4}{9}\lambda},
\ee
and one can obtain a blue curvaton power spectrum, $n_\sigma\simeq {\cal O}(2-4)$ \cite{stwo1}.

After inflation, the curvaton oscillates around the minimum of its potential and it decays into radiation. If we denote by $r$ the ratio of the curvaton energy density $\rho_\sigma$ over the radiation energy density $\rho_\gamma$,
 at small scales the linear curvature perturbation is dominated by the curvaton and becomes
 \begin{eqnarray}
 \zeta_1(\vx)&\simeq& r \zeta_{1,\sigma}(\vx)\,\,\,\,\,\,\,(k>k_c),\nonumber\\
r&=&\frac{3 \rho_\sigma}{4\rho_\gamma+3\rho_\sigma}.
 \end{eqnarray}
The perturbation is nevertheless getting a non-Gaussian contribution. Restricting ourselves to the quadratic potential (\ref{qq}), the curvature perturbation receives second-order contributions from the $(\delta\sigma(\vx))^2$ piece in $\delta\rho_\sigma$.  A standard computation leads to \cite{curvng,bmr,bmr2}
\be
\zeta_2(\vx)=r\zeta_{1,\sigma}(\vx)+r\left(\frac{3}{2}-r^2\right)\left(\zeta_{1,\sigma}(\vx)\right)^2= \zeta_1(\vx)+\frac{1}{r}\left(\frac{3}{2}-r^2\right)\left(\zeta_{1}(\vx)\right)^2,\,\,\,\,\,\,(k>k_c).
\ee
Neglecting for the moment the window functions, the corresponding skewness is
\be
S_3\simeq \frac{6}{r}\left(\frac{3}{2}-r^2\right),
\ee
which is a decreasing function of $r$ with  $S_3=3$ at $r=1$.
 Imposing again $S_3\lsim 0.06 \left(\sigma_{R_H}^2/0.02\right) \left(1.3/\zeta_c\right)^3$, one finds solutions only for sufficiently small thresholds or large variances. If we consider higher-order terms in the curvaton axion-like potential, higher-order cumulants are generated, but they will depend on other parameters, such as the quartic coupling $\sim m^2/\phi_*^2$ and the background value $\sigma_0$.  We do not expect cancellations to operate and make the Gaussian mass fraction of PBHs trustable.

To estimate the impact of non-Gaussianity, we can use alternatively the  density contrast (during the radiation phase) $\Delta(\vx)=(4/9a^2H^2)\nabla^2\zeta(\vx)$ for which $\Delta_c\simeq 2\sqrt{2}\zeta_c/9\simeq 0.4(\zeta_c/1.3)$.
The corresponding skewness reads
\begin{equation}
\begin{aligned}
S_3^\Delta(R_H)=6\cdot\left(\frac{4}{9}\right)^3\frac{1}{r\sigma_{\Delta}^4(R_H)}\left(\frac{3}{2}-r^2\right)\mathscr{S}^\Delta_3(k_*,R_H),
\end{aligned}\nonumber
\end{equation}
\begin{equation}
\begin{aligned}
\mathscr{S}^\Delta_3(k_*,R_H)=\int\frac{\d^3 k_1}{(2\pi)^3}\int\frac{\d^3 k_2}{(2\pi)^3}W(k_1,R_H)W(k_2,R_H)
W(|\vk_1+\vk_2|,R_H)
(k_1R_H)^2(k_2R_H)^2
\\
(|\vk_1+\vk_2|R_H)^2 P_\zeta(k_1) P_\zeta(k_2),
\end{aligned}\nonumber
\end{equation}
where $P_\zeta(k)=(2\pi^2/k^3)A^2(k_*)$ for $k>k_*$ and $P_\zeta(k)=(2\pi^2/k^3)A^2(k_*)(k/k_*)^{n_\sigma-1}$ for $k<k_*$ and
\be
\sigma_{\Delta}^2(R_H)=\left(\frac{4}{9}\right)^2\int\frac{\d^3 k}{(2\pi)^3}(kR_H)^4W^2(k,R_H) P_\zeta(k).
\ee
The numerical results are shown in Fig. 2 which makes  explicit that the Gaussian evaluation of the PBH abundance does not suffice. One obvious improvement will be to evaluate the one-loop variance induced by the non-Gaussianity
\be
\delta\sigma_{\Delta}^2(R_H)=2\left(\frac{4}{9}\right)^2 \frac{1}{r^2}\left(\frac{3}{2}-r^2\right)^2\int\frac{\d^3 k}{(2\pi)^3}(kR_H)^4W^2(k,R_H) \int\frac{\d^3 p}{(2\pi)^3}P_\zeta(p)P_\zeta(|\vk-\vec{p}|),
\ee
where the power spectra are of course the tree-level ones. This improvement will however require an improved computation of the  skewness and so on.
We again follow Ref. \cite{hu} and ask what is the impact of non-Gaussianity if we demand that the primordial PBH abundance corresponds to the
totality of the dark matter abundance today. Let us minimise the  influence of non-Gaussianity by setting $r=1$ and $S_3\simeq 3$. Under these circumstances, $\delta\sigma_{\Delta}^2(R_H)$ is still much smaller than the tree-level $\sigma_{\Delta}^2(R_H)$, and the Gaussian approach would give $\sigma_{\Delta}^2(R_H)\gsim 0.02$ \cite{hu}. This is in sharp contrast with our condition
(\ref{condition}) on $S_3$ which would deliver $S_3\lsim 0.06$.  A value of $S_3$ would deliver a dark matter abundance smaller by almost twenty orders of magnitude. This again indicates the sensitivity of the cosmological predictions to the non-Gaussianity.

\noindent
\begin{figure}[h!]
    \begin{center}
      \includegraphics[scale=.5,angle=360]{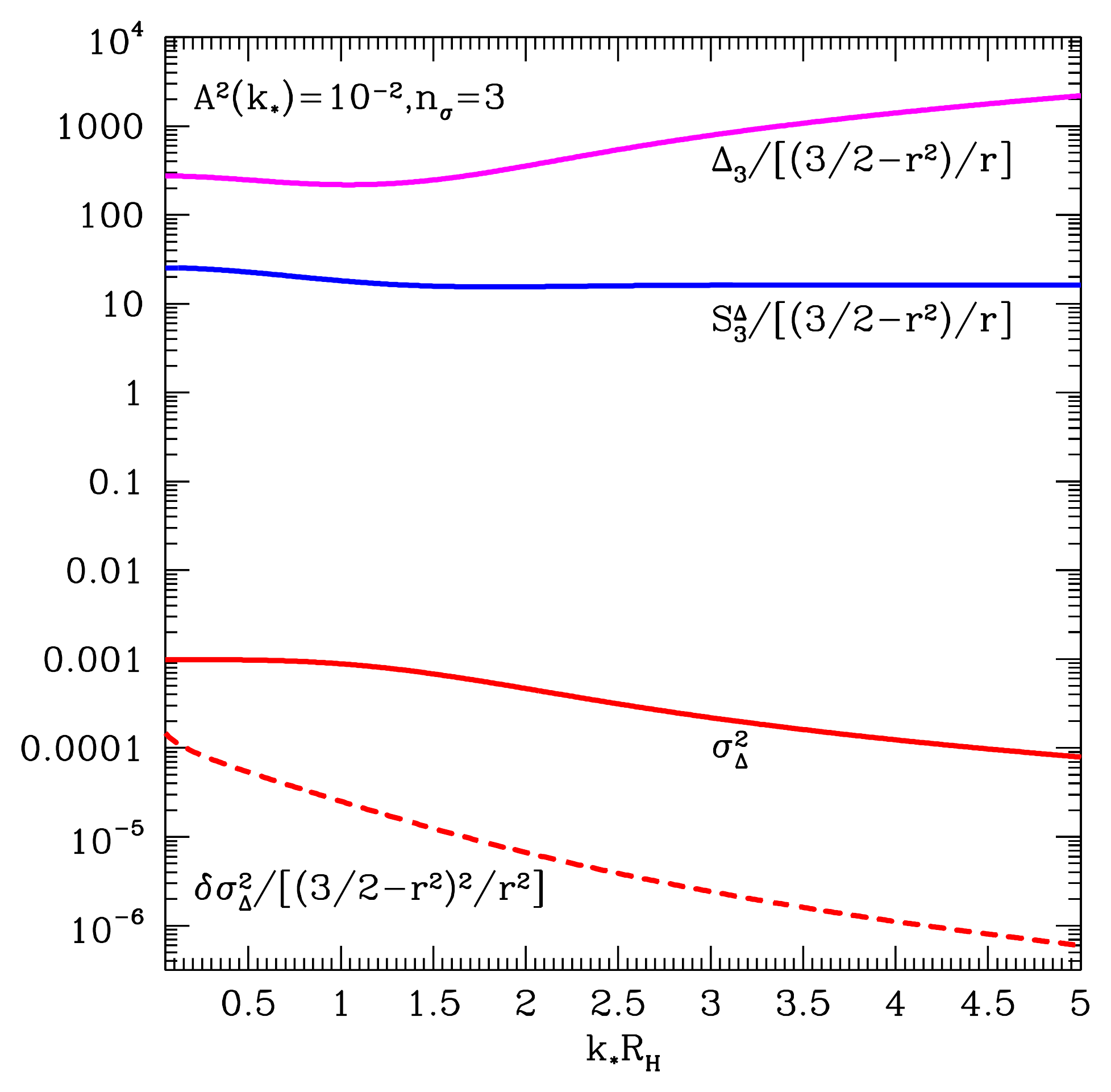}
    \end{center}
     \caption{\small The (normalized) tree-level and one-loop variances,  skewness and fine-tuning parameter of the density contrast $\Delta(\vx)$ as a function of $k_*R_H$ for $A^2(k_*)=10^{-2}$ and $n_\sigma=3$.}  
\end{figure}
\noindent 
\section{The clustering of PBHs in the presence of non-Gaussianity}
\noindent
The PBHs created in the radiation era and originated from spikes in the comoving curvature perturbation are highly clustered \cite{c}.
Using the peak theory model of bias, one can  compute the PBH $n$-point correlation functions. For Gaussian fluctuations one 
can follow Ref. \cite{pw} and show that the connected two-point correlator of PBHs is given by
\be\label{eq:rep}
\xi^{(2)}_{\nu,R_H}(\vx_1,\vx_2)=-1+\exp\left(\nu^2  w^{(2)}_{{R_H}}(\vx_1,\vx_2)\right).
\ee
One can easily see that the correlator $\xi^{(2)}_{\nu,R_H}(\vx_1,\vx_2)$ increments for higher values of $\nu$. In other words, since PBHs are generated by large  fluctuations and their mass fraction is exponentially suppressed for $\nu\gg 1$, the fewer there are of them, the more clustered they are.

The technique explained in Ref. \cite{blm}  allows also to compute the correlators of the density contrasts above a given threshold for a non-Gaussian
statistics.
Let us for instance  consider the case of $W(\vx,R)=\delta^{(3)}(\vx)$. It follows immediately that in the large threshold limit the joint probability is
\begin{eqnarray}
\Pi_{\nu,R}^{(N)}(\vx_1,\cdots,\vx_N)=\frac{e^{-N \nu^2}}{(2\pi \nu^2)^{N/2}}
\prod_{i=1}^N Z[J_i],
\end{eqnarray}
which, by using Eq. (\ref{final}) can be written as 
\begin{eqnarray}
\Pi_{\nu,R}^{(N)}(\vx_1,\cdots,\vx_N)=\langle \rho_{\nu,R}(\vx)\rangle^N
\exp\left\{\sum_{n=2}^\infty \frac{\nu^n}{n!} \bigg[ \sum_{i_1,\cdots,i_n=1}^N w^{(n)}_{{R},[i_n]}-N w^{(n)}_R\bigg]\right\}.
\end{eqnarray}
Therefore, the $N$-point peak disconnected correlation functions $\xi^{(N)}_{{\rm disc},\nu,R}(\vx_1,\cdots \vx_N)$ defined as 
\begin{eqnarray}
\xi^{(N)}_{{\rm disc},\nu,R}(\vx_1,\cdots, \vx_N)=\left\langle \prod_{i=1}^N\frac{\rho_{\nu,R}(\vx_i)}{\langle\rho_{\nu,R}\rangle}\right\rangle -1,
\end{eqnarray}
turn out to be 
\begin{eqnarray}
\xi^{(N)}_{{\rm disc},\nu,R}(\vx_1,\cdots, \vx_N) =-1+\exp\left\{\sum_{n=2}^\infty \frac{\nu^n}{n!} \bigg[ \sum_{i_1,\cdots,i_n=1}^N w^{(n)}_{{R},[i_n]}-N w^{(n)}_R\bigg]\right\}.
\end{eqnarray}
For instance, the non-Gaussian  two and three--point peak connected correlation functions 
read respectively
\begin{eqnarray}
\xi^{(2)}_{\nu,R}(\vx_1,\vx_2)&=&-1+\exp\left\{\sum_{n=2}^\infty\sum_{j=1}^{n-1} \frac{\nu^n}{j!(n-1)!}  w^{(n)}_{{R},[j;n-j]}\bigg]\right\}\nonumber\\
&=&-1+\exp\left\{\sum_{n=2}^\infty\sum_{j=1}^{n-1} \frac{\nu^n\sigma_{R}^{-n}}{j!(n-1)!}  \xi^{(n)}_{{R}}
\left(\underset{j-{\rm times}}{\vx_1,\cdots,\vx_1},\underset{(n-j)-{\rm times}}{\vx_2,\cdots,\vx_2}
\right)\right\},
\end{eqnarray}
and
\begin{equation}
\begin{aligned}
\xi^{(3)}_{\nu,R}(\vx_1,\vx_2,\vx_3)=F_{\nu,R}(\vx_1,\vx_2,\vx_3)\left[\left(\xi^{(2)}_{\nu,R}(\vx_1,\vx_2)\xi^{(2)}_{\nu,R}(\vx_2,\vx_3)+{\rm cyclic}\right)+\xi^{(2)}_{\nu,R}(\vx_1,\vx_2)\xi^{(2)}_{\nu,R}(\vx_2,\vx_3)\xi^{(2)}_{\nu,R}(\vx_3,\vx_1)\right]
\\
+\left[F_{\nu,R}(\vx_1,\vx_2,\vx_3)-1\right]\left[1+\xi^{(2)}_{\nu,R}(\vx_1,\vx_2)+\xi^{(2)}_{\nu,R}(\vx_2,\vx_3)+\xi^{(2)}_{\nu,R}(\vx_3,\vx_1)\right],
\end{aligned}
	\end{equation}
with
\be
F_{\nu,R}(\vx_1,\vx_2,\vx_3)=\exp\left\{\sum_{n=3}^\infty\sum_{j=1}^{n-2}\sum_{k=1}^{n-j-1} \frac{(\zeta_c/\sigma^2_{R})^n}{j!(n-1)!}  \xi^{(n)}_{{R},[j;k; n-j-k]}\bigg]\right\}.
\ee
As already stressed in Ref. \cite{blm} (see also Ref. \cite{gw}) for the validity of these equations,  correlations do not need to  be small either with respect to unity or with respect to the variance on the smoothing scale. The two-point correlator can be rewritten
as
\begin{eqnarray}
\xi^{(2)}_{\nu,R_H}(\vx_1,\vx_2)&=&-1+\exp\left(\nu^2/\sigma^2_{R_H} \xi^{(2)}_{{R_H}}(\vx_1,\vx_2)\right)\exp\left(\nu^3/\sigma^3_{R_H} \xi^{(3)}_{{R_H}}(\vx_1,\vx_2,\vx_2)+\cdots\right).
\end{eqnarray}
The power spectrum in momentum space and over the normalization volume $V$ will read
\begin{equation}
P_{\nu,R_H}(k)=\frac{4\pi}{V}\int_{R_H}^\infty\d r\,r^2\,\frac{\sin k r}{kr}\left[\exp\left(\nu^2/\sigma^2_{R_H} \xi^{(2)}_{{R_H}}(r)\right)\exp\left(\nu^3/\sigma^3_{R_H} \xi^{(3)}_{{R_H}}(r)+\cdots\right)-1\right],\,\,\,\, r=|\vx_1-\vx_2|,
\end{equation}
where the lower cut-off of the integral is due to the 
finite size of the PBHs. One can expand in powers the exponentials and obtain
\be
P_{\nu,R_H}(k)=\frac{4\pi}{V}\int_{R_H}^\infty\d r\,r^2\,\frac{\sin k r}{kr}\left[\frac{\nu^2}{\sigma^2_{R_H}} \xi^{(2)}_{{R_H}}(r)+\frac{\nu^3}{\sigma^3_{R_H}} \xi^{(3)}_{{R_H}}(r)+\cdots\right],
\ee
For instance, for the class of single-field model generating PBHs at inflection points  described in section IV one finds that the Fourier transform of
$\nu^3/\sigma^3_{R_H} \xi^{(3)}_{{R_H}}(\vx_1,\vx_2,\vx_2)$ is given by
\begin{equation}
\begin{aligned}
3\frac{\nu^3}{\sigma^3_{R_H}} 
W_{R_H}(k)P_\zeta(k)\int \frac{\d^3 q}{(2\pi)^3}\,  \,W_{R_H}(q)\,P_\zeta(q)\,W_{R_H}(|\vq+\vk|)\left[
\frac{P_\zeta(|\vq+\vk|)}{P_\zeta(k)}+2\right]\approx 6\frac{\nu^3}{\sigma^3_{R_H}} \sigma^2_{R_H}W_{R_H}(k)P_\zeta(k)\\
=6\,\zeta_c\frac{\nu^2}{\sigma^2_{R_H}} W_{R_H}(k)P_\zeta(k),
\end{aligned}
	\end{equation}
that is a factor $\sim 6\zeta_c$ times the Gaussian contribution.
One sees again that the non-Gaussian corrections can alter the clustering properties significantly. Since  clustering influences the merging of PBHs, this could have a severe impact on the subsequent cosmological evolution of PBHs. For instance by changing the limits on initial PBH abundance or, as 
  merging enhances   the abundance of more massive and  long-lived PBHs, by making PBHs  the  seeds of supermassive BHs \cite{ch2,Bernal}. Our preliminary analysis above suggests that 
  non-Gaussianity may not be disregarded in these considerations.
We will defer the study of this subject in a forthcoming
publication.

\section{Conclusions}
\noindent
The idea that PBHs may be  the dark matter in our universe is a fascinating one. Equally  exciting is the fact that PBHs may owe their origin to 
 large and rare fluctuations generated during inflation, i.e. during the phase of accelerated expansion which might be  ultimately responsible also for the large-scale structures we see around us. If this is the mechanism nature has chosen to permeate the universe with dark matter, it is
 vital to understand in details the quantitative predictions within a given set-up. In this paper we have discussed the role of non-Gaussianity in determining
 the PBH mass fraction at formation time and their clustering properties. We have done so by discussing a path-integral procedure which leads to exact
 expressions for the quantities of interest and by estimating the impact of non-Gaussianity in a couple of models.
 Our findings show that non-Gaussianity  should be taken  into account when considering a model for PBH production and the corresponding non-Gaussianity parameters should be evaluated in order to gauge if PBHs are interesting candidates for dark matter and when analysing their clustering properties and subsequent merging and accretion. Also, it will be interesting to investigate
 the impact of the non-Gaussianity on the production of the gravity waves from inflation generated by the same large perturbations originating the PBHs \cite{revPBH} and on
 the limits coming from CMB $\mu$-distortions \cite{mu}.

\bigskip
\bigskip
\noindent
\subsection*{Acknowledgments}
\noindent
 S.M.  acknowledges partial financial support by the ASI/INAF Agreement 2014-024-R.0 for the Planck LFI Activity of Phase E2. A.R. is  supported by the Swiss National Science Foundation (SNSF), project {\sl Investigating the Nature of Dark Matter}, project number: 200020-159223. A.K. thanks the Cosmology group at the D\'epartement de Physique 
  Th\'eorique  at the  Universit\'e de Gen\`eve for the kind hospitality and financial support. G.F. and A.R. also thank the Galileo Galilei Institute for Theoretical Physics, Florence (Italy), where this work was finalised, for the nice hospitality. 




\begin{thebibliography}{99}
  \bibitem{ligo} B. P. Abbott et al. [LIGO Scientific and Virgo Collaborations], Phys. Rev. Lett. 116, 061102 (2016)
\arXiv{1602.03837}{gr-qc}.
  
  \bibitem{PBH1} B. J. Carr and S. W. Hawking, Mon. Not. Roy. Astron.
Soc. {\bf 168}, 399 (1974).
  
  \bibitem{PBH2}  P. Meszaros, Astron. Astrophys. {\bf 37}, 225 (1974).
  
  \bibitem{PBH3} B. J. Carr, Astrophys. J. {\bf 201}, 1 (1975).
  
  \bibitem{kam} S.~Bird, I.~Cholis, J.~B.~MuÃ±oz, Y.~Ali-HaÃ¯moud, M.~Kamionkowski, E.~D.~Kovetz, A.~Raccanelli and A.~G.~Riess,
  Phys.\ Rev.\ Lett.\  {\bf 116}, no. 20, 201301 (2016)
  \arXiv{1603.00464}{astro-ph.CO}.
  
\bibitem{rep1} 
  S.~Clesse and J.~García-Bellido,
  Phys.\ Dark Univ.\  {\bf 15}, 142 (2017)
  \arXiv{1603.05234}{astro-ph.CO}.
 
\bibitem{rep2} 
  M.~Sasaki, T.~Suyama, T.~Tanaka and S.~Yokoyama,
  Phys.\ Rev.\ Lett.\  {\bf 117}, no. 6, 061101 (2016)
  \arXiv{1603.08338}{astro-ph.CO}.
  
  
  \bibitem{revPBH} 
  M.~Sasaki, T.~Suyama, T.~Tanaka and S.~Yokoyama,
  \arXiv{1801.05235}{astro-ph.CO}.
 
 \bibitem{c0} B.~Carr, M.~Raidal, T.~Tenkanen, V.~Vaskonen and H.~Veerme,
  Phys.\ Rev.\ D {\bf 96}, no. 2, 023514 (2017)
   \arXiv{1705.05567}{astro-ph.CO}.
 
  \bibitem{c1} M.~Zumalacarregui and U.~Seljak,
  \arXiv{1712.02240}{astro-ph.CO}.
  
  \bibitem{c2} 
  J.~Garcia-Bellido, S.~Clesse and P.~Fleury,
 \arXiv{1712.06574}{astro-ph.CO}
 
    \bibitem{rev} S.~Clesse and J.~Garcia-Bellido,
  \arXiv{1711.10458}{astro-ph.CO}.

\bibitem{carrsilk}  B.~Carr and J.~Silk,
  \arXiv{1801.00672}{astro-ph.CO}.
 

  
\bibitem{qcd} K. Jedamzik,  Phys. Rev. D {\bf 55}, 5871 (1997)
\arXivold{astro-ph/9605152}.
  
\bibitem{bqcd} C.~T.~Byrnes, M.~Hindmarsh, S.~Young and M.~R.~S.~Hawkins,
  \arXiv{1801.06138}{astro-ph.CO}.  
  
  \bibitem{pt} K. Jedamzik and J. C. Niemeyer, 
Phys. Rev. D {\bf 59}, 124014 (1999) \arXivold{astro-ph/9901293}.
  
  \bibitem{s1} P.~Ivanov, P.~Naselsky and I.~Novikov,
  Phys.\ Rev.\ D {\bf 50}, 7173 (1994).
  
  \bibitem{s2} J.~Garc\'{\i}a-Bellido, A.D.~Linde and D.~Wands,
  Phys.\ Rev.\ D {\bf 54} (1996) 6040
  \arXivold{astro-ph/9605094}.

  \bibitem{s3} 
  P.~Ivanov, Phys.\ Rev.\ D {\bf 57}, 7145 (1998)
  \arXivold{astro-ph/9708224}.
    
    
    \bibitem{sone1} J.~Garcia-Bellido and E.~Ruiz Morales,
  Phys.\ Dark Univ.\  {\bf 18}, 47 (2017)
   \arXiv{1702.03901}{astro-ph.CO}.
    
    \bibitem{sone0} K.~Kannike, L.~Marzola, M.~Raidal and H.~Veerme,
  JCAP {\bf 1709}, no. 09, 020 (2017)
  \arXiv{1705.06225}{astro-ph.CO}.

  
\bibitem{sone2}G.~Ballesteros and M.~Taoso,
  Phys.\ Rev.\ D {\bf 97}, no. 2, 023501 (2018)
  \arXiv{1709.05565}{hep-ph}.
    
    
  \bibitem{stwo1} M.~Kawasaki, N.~Kitajima and T.~T.~Yanagida,
  Phys.\ Rev.\ D {\bf 87}, no. 6, 063519 (2013)
  \arXiv{1207.2550}{hep-ph}.

  
      \bibitem{Carr} B.~Carr, F.~Kuhnel and M.~Sandstad,
  Phys.\ Rev.\ D {\bf 94}, 083504 (2016)
  \arXiv{1607.06077}{astro-ph.CO}.

 \bibitem{gauge} J. Garcia-Bellido, M. Peloso and C. Unal, JCAP 1612, no. {\bf 12}, 031 (2016) 
 \arXiv{1610.03763}{astro-ph.CO}.
 
 \bibitem{cs} B.~Carr, T.~Tenkanen and V.~Vaskonen,
  Phys.\ Rev.\ D {\bf 96}, no. 6, 063507 (2017)
  \arXiv{1706.03746}{astro-ph.CO}.
  
    \bibitem{ssm1}
  J.M.~Ezquiaga, J.~Garc\'{\i}a-Bellido and E.~Ruiz Morales,
  \arXiv{1705.04861}{astro-ph.CO}.
  
  \bibitem{ssm2} J.~R.~Espinosa, D.~Racco and A.~Riotto,
\arXiv{1710.11196}{hep-ph}.
  
 \bibitem{harada} T. Harada, C. M. Yoo and K. Kohri, Phys. Rev. D {\bf 88}, 084051 (2013), Erratum: [Phys. Rev. D {\bf 89}, 029903 (2014)],
\arXiv{1309.4201}{astro-ph.CO}.
 
 \bibitem{bb} S. Young, C. T. Byrnes, and M. Sasaki, JCAP {\bf 1407}, 045 (2014)
 \arXiv{1405.7023}{gr-qc}.
 
 \bibitem{miller} I. Musco and J. C. Miller, Class. Quant. Grav. {\bf 30}, 145009 (2013) \arXiv{1201.2379}{gr-qc}.
 
 \bibitem{hu} H.~Motohashi and W.~Hu,
  Phys.\ Rev.\ D {\bf 96}, no. 6, 063503 (2017)
  \arXiv{1706.06784}{astro-ph.CO}.
   

 \bibitem{ls} A.~M.~Green, A.~R.~Liddle, K.~A.~Malik and M.~Sasaki,
  Phys.\ Rev.\ D {\bf 70}, 041502 (2004)
  \arXivold{astro-ph/0403181}.

   \bibitem{c} J.~R.~Chisholm,
  Phys.\ Rev.\ D {\bf 74}  (2006) 043512
  \arXivold{astro-ph/0604174}.
    
      \bibitem{jn} K.~Jedamzik and J.~C.~Niemeyer,
  Phys.\ Rev.\ D {\bf 59} (1999), 124014
  \arXivold{astro-ph/9901293}.
     
        

 



 \bibitem{cg} S.~Clesse and J.~Garc\'{\i}a-Bellido,
  Phys.\ Rev.\ D {\bf 92} (2015) 023524
  \arXiv{1501.07565}{astro-ph.CO}. 
  
  
  \bibitem{asp} F.~KÃŒhnel and M.~Sandstad,
  Phys.\ Rev.\ D {\bf 94}, no. 6, 063514 (2016)
  \arXiv{1602.04815}{astro-ph.CO}.

  \bibitem{misao} M.~Shibata and M.~Sasaki,
  Phys.\ Rev.\ D {\bf 60}, 084002 (1999)
    \arXivold{gr-qc/9905064}.


  
  \bibitem{cc} B. Carr, K. Kohri, Y. Sendouda and J. Yokoyama,
  Phys.\ Rev.\ D {\bf 81} (2010) 104019
  \arXiv{0912.5297}{astro-ph.CO}.

\bibitem{byrnes} S.~Young and C.~T.~Byrnes,
  JCAP {\bf 1308}, 052 (2013)
  \arXiv{1307.4995}{astro-ph.CO}.

\bibitem{ngtwo} E.~V.~Bugaev and P.~A.~Klimai,
  Int.\ J.\ Mod.\ Phys.\ D {\bf 22}, 1350034 (2013)
  \arXiv{1303.3146}{astro-ph.CO}.
  
   \bibitem{ng1} J. S. Bullock and J. R. Primack, Phys. Rev. {\bf D}55, 7423 (1997), 
  \arXivold{astro-ph/9611106}.
  
  \bibitem{ng2} J. Yokoyama, Phys. Rev. {\bf D}58, 083510 (1998), 
  \arXivold{astro-ph/9802357}.
  
  \bibitem{ng3} R. Saito, J. Yokoyama, and R. Nagata, JCAP {\bf 0806}, 024 (2008), \arXiv{0804.3470}{astro-ph.CO}.
  
  \bibitem{ng33} C.~T.~Byrnes, E.~J.~Copeland and A.~M.~Green,
  Phys.\ Rev.\ D {\bf 86}, 043512 (2012)
 \arXiv{1206.4188}{astro-ph.CO}.
 
  \bibitem{ng4} S. Young, D. Regan, and C. T. Byrnes, JCAP {\bf 81602}, 029 (2016), \arXiv{1512.07224}{astro-ph.CO}.
  
  \bibitem{ng5} M. Kawasaki and Y. Tada, JCAP {\bf 1608}, 041 (2016), \arXiv{1512.03515}{astro-ph.CO}.
  
  \bibitem{ng6} C. Pattison, V. Vennin, H. Assadullahi, and D. Wands, JCAP {\bf 1710}, 046 (2017),
\arXiv{1707.00537}{hep-th}.
  
  
 \bibitem{planck} P.~A.~R.~Ade {\it et al.} [Planck Collaboration],
  Astron.\ Astrophys.\  {\bf 594}, A17 (2016)
  \arXiv{1502.01592}{astro-ph.CO}.
  
 \bibitem{localNG} Y.~Tada and S.~Yokoyama,
  Phys.\ Rev.\ D {\bf 91}, no. 12, 123534 (2015)
  \arXiv{1502.01124}{astro-ph.CO}.

  \bibitem{byrnes2} S.~Young and C.~T.~Byrnes,
  JCAP {\bf 1504}, no. 04, 034 (2015)
  \arXiv{1503.01505}{astro-ph.CO}.
  
 
  \bibitem{second} K.~A.~Malik and D.~Wands,
  Class.\ Quant.\ Grav.\  {\bf 21}, L65 (2004)
 \arXivold{astro-ph/0307055}.
 
  
  \bibitem{blm} S.~Matarrese, F.~Lucchin and S.~A.~Bonometto,
  Astrophys.\ J.\  {\bf 310}, L21 (1986).
   
   \bibitem{gw} B.~Grinstein and M.~B.~Wise,
  Astrophys.\ J.\  {\bf 310}, 19 (1986).
   
 \bibitem{bg}  R.~Barbieri and G.~F.~Giudice,
  Nucl.\ Phys.\ B {\bf 306}, 63 (1988).
  
  \bibitem{lrreview}   D.H.~Lyth and A.~Riotto,
  Phys.\ Rept.\  {\bf 314} (1999) 1
  \arXivold{hep-ph/9807278}.
  
     
   
   
   \bibitem{u1} W. H. Kinney, Phys. Rev. D {\bf 72}, 023515 (2005)
\arXivold{gr-qc/0503017}.

\bibitem{u2} C. Dvorkin and W. Hu, Phys. Rev. {\bf D} 81, 023518 (2010) \arXiv{0910.2237}{astro-ph.CO}.

\bibitem{u3} M. H. Namjoo, H. Firouzjahi, and M. Sasaki, Europhys. Lett. {\bf 101}, 39001 (2013) \arXiv{1210.3692}{astro-ph.CO}.

\bibitem{u4}X.~Chen, H.~Firouzjahi, M.~H.~Namjoo and M.~Sasaki,
  EPL {\bf 102}, no. 5, 59001 (2013)
   \arXiv{1301.5699}{hep-th}.

\bibitem{u5} C.~Germani and T.~Prokopec,
  Phys.\ Dark Univ.\  {\bf 18}, 6 (2017)
  \arXiv{1706.04226}{astro-ph.CO}.
  
\bibitem{u6} K.~Dimopoulos,
  Phys.\ Lett.\ B {\bf 775}, 262 (2017)
  \arXiv{1707.05644}{hep-ph}.


\bibitem{pajer} B.~Finelli, G.~Goon, E.~Pajer and L.~Santoni,
  \arXiv{1711.03737}{hep-th}.

  \bibitem{rev1}  Y.~F.~Cai, X.~Chen, M.~H.~Namjoo, M.~Sasaki, D.~G.~Wang and Z.~Wang,
  \arXiv{1712.09998}{astro-ph.CO}.
 
 \bibitem{diagrammatic} C.~T.~Byrnes, K.~Koyama, M.~Sasaki and D.~Wands,
  JCAP {\bf 0711}, 027 (2007)
   \arXiv{0705.4096}{hep-th}.
   
   
\bibitem{rep3} 
  A.~A.~Starobinsky,
  Phys.\ Lett.\  {\bf 117B}, 175 (1982).
  
\bibitem{rep4} 
  A.~A.~Starobinsky,
  JETP Lett.\  {\bf 42}, 152 (1985)
  [Pisma Zh.\ Eksp.\ Teor.\ Fiz.\  {\bf 42}, 124 (1985)].
  
\bibitem{deltaN} M.~Sasaki and E.~D.~Stewart,
  Prog.\ Theor.\ Phys.\  {\bf 95}, 71 (1996)
   \arXivold{astro-ph/9507001}.
   

  \bibitem{curvaton} D.H.~Lyth and D.~Wands,
  Phys.\ Lett.\ B {\bf 524} (2002) 5
  \arXivold{hep-ph/0110002}.  


\bibitem{curvng} D.~H.~Lyth, C.~Ungarelli and D.~Wands,
  Phys.\ Rev.\ D {\bf 67}, 023503 (2003)
  \arXivold{astro-ph/0208055}.


\bibitem{bmr} N.~Bartolo, S.~Matarrese and A.~Riotto,
  Phys.\ Rev.\ D {\bf 69}, 043503 (2004)
  \arXivold{hep-ph/0309033}.
  
  \bibitem{bmr2}  N.~Bartolo, S.~Matarrese and A.~Riotto,
  JCAP {\bf 0401}, 003 (2004)
  \arXivold{astro-ph/0309692}.
  
  \bibitem{pw} H.~D.~Politzer and M.~B.~Wise,
  Astrophys.\ J.\  {\bf 285}, L1 (1984).
    
  
  \bibitem{ch2} J.~R. Chisholm,
  Phys.\ Rev.\ D {\bf 84}, 124031 (2011)
  \arXiv{1110.4402}{astro-ph.CO}.
  
  \bibitem{Bernal} 
  J.~L.~Bernal, A.~Raccanelli, L.~Verde and J.~Silk,
  \arXiv{1712.01311}{astro-ph.CO}.
  
  \bibitem{mu} T.~Nakama, B.~Carr and J.~Silk,
  \arXiv{1710.06945}{astro-ph.CO}.

   \end{thebibliography}
\end{document}